# Wellbeing-Centered UX: Supporting Content Moderators

[Diana Mihalache](#) and [Dalila Szostak](#)

## Abstract

This chapter focuses on the intersection of user experience (UX) and wellbeing in the context of content moderation. Human content moderators play a key role in protecting end users from harm by detecting, evaluating, and addressing content that may violate laws or product policies. They face numerous challenges, including exposure to sensitive content, monotonous tasks, and complex decisions, which are often exacerbated by inadequate tools. This chapter explains the importance of incorporating wellbeing considerations throughout the product development lifecycle, offering a framework and practical strategies for implementation across key UX disciplines: research, writing, and design. By examining these considerations, this chapter provides a roadmap for creating user experiences that support content moderators, benefiting both the user and the business.

**Wellbeing-Centered UX: Supporting Content Moderators**

Human content moderators play a crucial role in ensuring the quality, accuracy, and appropriateness of digital content, and yet their work remains largely invisible to the public (Gillespie, 2018; Lenaerts & Waeyaert, 2022; Steiger et al., 2021). Human content moderators are tasked with filtering through vast amounts of user-generated content (UGC) and making thousands of complex decisions on what violates platform policies or laws (e.g., depictions of violence, abuse, hate speech, spam, or other illegal material). Despite automation's assistance in flagging potentially violating content, human involvement remains indispensable due to the sophistication required for nuanced judgments, the high cost of errors, and the ever-changing landscape of content and policies. The inherent complexity of this decision-making, coupled with varying levels of exposure to egregious content, can impact human moderator wellbeing. Although workplace wellness is often considered the domain of benefits or operations teams, user experience (UX) can also play an important role in protecting the wellbeing of content moderators.

**What is User Experience?**

Before diving into the world of content moderation and UX, it is important to define it. User experience (UX) is a multifaceted concept centered on a person's perceptions and responses that result from the use or anticipated use of a product, system, or service. The International Organization for Standardization defines UX as encompassing "all the users' emotions, beliefs, preferences, perceptions, physical and psychological responses, behaviors and accomplishments that occur before, during and after use" (ISO, 2019). This definition highlights the holistic and subjective nature of the interaction, extending beyond mere task completion to include the user's entire affective and cognitive journey.

Don Norman, who is credited with coining the term, emphasizes that user experience encompasses all aspects of the end-user's interaction with the company, its services, and its products (Norman & Nielsen, n.d.). This perspective moves beyond the digital interface to consider the entire ecosystem of touchpoints that shape a user's perception. Further academic models have deconstructed UX into constituent components. For instance, Hassenzahl and Tractinsky (2006) differentiate between pragmatic qualities, which relate to traditional usability and the achievement of "do-goals," and hedonic qualities, which relate to the fulfillment of psychological needs such as stimulation, identity, and relatedness. Similarly, Morville (2004) proposes a "honeycomb" framework, identifying seven facets that contribute to a valuable user experience: usefulness, usability, desirability, findability, accessibility, credibility, and value. Collectively, these frameworks establish UX as a dynamic, context-dependent phenomenon that is crucial for understanding the interaction between humans and the products, systems, and tools used to complete any given task.

In the context of Trust and Safety, User Experience (UX) is the overall quality of interaction that people have with the platform or product's rules, reporting systems, and enforcement processes. Beyond product features and their usability, it is about the human side of policies. This includes how clear and straightforward it is for a person to report harmful content, how fair and understandable an appeals process feels to a creator whose content was removed, and how efficient and sustainable the review tools are for content moderators. In essence, UX is the practical application of Trust and Safety principles, directly shaping how the community perceives the platform or product as fair, trustworthy, and safe, which in turn drives better compliance and higher-quality reporting.

**Challenges to Wellbeing in Content Moderation**

A content moderator is an essential, albeit often invisible, contributor to the health of the online world. Their primary role is to review user-generated content, which can include text, images, videos, and audio, to ensure it complies with a platform's policies, community guidelines, and legal standards. Content moderation is widely recognized as necessary to establish a safe, respectful, and trustworthy online environment, and that it presents notable challenges for those who do the work (Ahmad, 2019; Ahmad & Krzywdzinski, 2022; Dang et al., 2018; Gillespie, 2018; Steiger et al., 2021). The work of a content moderator is characterized by a unique combination of stressors: it is highly repetitive, involves exposure to sensitive content, and demands rapid complex, high-stakes decisions. While direct research into the specific impact of content moderation on wellbeing is still nascent, a broader body of literature on these individual occupational hazards provides a critical framework for understanding this role.

*Task Repetition*

The work of a content moderator is fundamentally defined by relentless task repetition, performed at scale. The work is largely characterized by highly repetitive tasks with short cycle times, a structure often likened to call centers or assembly line work (Bharucha et al., 2022; Lenaerts & Waeyaert, 2022; Ruckenstein & Turunen, 2020; Steiger et al., 2021). It can be described as a high-demand, low-autonomy environment, which is explained by the demand-control model (Isaksen 2000; Lundberg et al., 1989; Weber et al., 1980). The model posits that work with high psychological demands paired with low decision latitude is linked to adverse psychological outcomes, including an increased risk of depression and loss of self-esteem (LaMontagne et al., 2014; Oeij et al., 2017).

On a psychological level, such repetitive tasks frequently induce monotony and boredom, particularly in the absence of novel stimuli (Krueger, 1989). Physiologically, highly repetitive tasks are sometimes associated with elevated heart rate, increased adrenaline excretion, and muscle tension (Weber et al., 1980). For example, studies of assembly line workers show that hormones like catecholamines correlate with time pressures and demands, and cortisol is linked to irritation, tenseness, and tiredness (Lundberg et al., 1989). Tying these factors together, a key driver of stress in these roles is a perceived lack of influence and autonomy (Briner, 2000; Manganelli et al., 2018; Richardson, 2017; Scheuerman et al., 2021; Slemp et al., 2015; Steiger et al., 2021).

### *Exposure to Sensitive Content*

A significant challenge for content moderators is the repeated and prolonged exposure to sensitive, and egregious content, such as violence, child abuse, and hate speech (Ahmad & Krzywdzinski, 2022; Bharucha et al., 2022; Lenaerts & Waeyaert, 2022; Scheuerman et al., 2021; Vîrgă et al., 2020). This can place content moderators at a higher risk for experiencing psychological distress and conditions like post-traumatic stress disorder (PTSD), vicarious trauma, anxiety, and depression (Lenaerts & Waeyaert, 2022; Steiger et al., 2021). Potential symptoms include intrusive thoughts, emotional numbing, avoidance behaviors, and shifts in worldview, particularly concerning perceptions of personal safety (Antony et al., 2020; Barrett, 2020; LaMontagne et al., 2014; Newell & MacNeil, 2010; Steinlin et al., 2017). Factors such as limited workplace support, stringent performance metrics, and non-disclosure agreements (NDAs) that restrict discussion about their experiences can exacerbate these risks, potentially leading to isolation (Lenaerts & Waeyaert, 2022).

As longitudinal studies in exposure to sensitive content for moderators are limited at the moment, studies into adjacent professions with regular exposure to traumatic material or survivor narratives—such as emergency dispatchers, journalists, child welfare professionals, and social workers—offer valuable insights (Baird & Jenkins, 2003; Barleycorn, 2022; Berget et al., 2012; Castanheira & Chambel, 2010; Caringi et al., 2012, 2017; Handa et al., 2010; Lanza et al., 2018; Steinlin et al., 2017; Trippany et al., 2011; Wild et al., 2016). These studies show emotional and psychological consequences, including heightened risks for conditions like secondary traumatic stress, compassion fatigue, and PTSD symptoms. Furthermore, we can also look to the protective factors and interventions developed in these fields to inform strategies for supporting content moderators.

### *Complex Decisions*

As automation increasingly handles straightforward moderation cases, human content moderators are predominantly tasked with navigating complex "gray area" content: material that does not explicitly violate a platform's rules but sits in an ambiguous space that can still cause harm or be disruptive (Scheuerman et al., 202l; Steiger et al, 2021). This work demands high decision complexity and sensitivity, requiring content moderators to interpret nuanced policies, cultural contexts, and country-specific guidelines (Ahmad & Krzywdzinski, 2022). Successfully managing these complex decisions takes significant cognitive effort while sometimes also requiring emotional bandwidth when moderating sensitive content (Bharucha, 2022; Lenaerts & Waeyaert, 2022; Scheuerman et al., 202l).

This is important to highlight as cognitive processes like executive functioning and emotion regulation are intimately linked (Caringi et al., 2012; Gyurak et al., 2012; Krueger, 1989; Weber et al., 1989). When one is emotionally dysregulated, it is more difficult to manage cognitive load, and vice versa, higher cognitive load makes it more difficult to regulate emotions. These challenges are further exacerbated by fragmented tooling ecosystems that necessitate repeated switching between multiple tools and policy resources (Bharucha, 2022). Consequently, well designed tools and a strong user experience are far more critical to content moderator wellbeing than one might initially assume.

## Conceptualizing Wellbeing: A Multidimensional Perspective

Before describing specific characteristics of wellbeing for content moderators, it is helpful to take a step back and look at various models of wellbeing from a multidisciplinary perspective. Academic literature suggests that while terms like wellness, wellbeing, quality of life (QoL), and life satisfaction are often used interchangeably, there is no single, unified conceptual framework (Miller, 2005; Miller & Foster, 2010). Generally, wellbeing is described as a dynamic and multifaceted concept that has broadened from simply the absence of disease to encompass a positive and holistic view. Different theoretical frameworks conceptualize wellbeing through hedonic approaches, which emphasize pleasure and happiness, such as subjective wellbeing (Diener et al., 1985; 1997; Ryan & Deci, 2001), and eudaimonic approaches, which focus on fulfilling one's potential and optimal functioning (Lent, 2004; Palombi, 1992; Ryan & Deci, 2001; Ryff, 1989; Ryff & Keyes, 1995). Additionally, QoL models include broader physical, psychological, and social dimensions, which are integrated into optimizing functioning (Cooke et al., 2016). Broadly models of wellbeing often include physical, psychological, social, intellectual, spiritual, occupational, and environmental aspects. While specific frameworks and definitions vary, they all stress a multidimensional conceptualization of wellbeing and advocate for a holistic view with interconnected dimensions.

For the purpose of this chapter, we distill these concepts into three interdependent attributes critical for wellbeing in the context of designing tools for content moderators: (1) enabling effectiveness within the role, (2) fostering connection, and (3) cultivating resilience. Importantly, we highlight that reactive intervention is not sufficient for wellbeing, and requires a holistic approach (Cooke et al., 2016; LaMontagne et al., 2014; Wassell & Dodge, 2015). A multi-pronged strategy should begin with prevention by identifying and mitigating potential stressors well in advance (LaMontagne et al., 2014; Steiger, 2021). When prevention is not possible, intervention is most helpful when it is timely, offering support during or immediately after adverse events (Bryant, 2021; Caringi et al., 2012). Lastly, wellbeing extends beyond the mere absence of negative effects, and a comprehensive wellbeing strategy should involve cultivating positive experiences (Cooke et al., 2016; Miller & Foster, 2010; Richardson & Rothstein, 2008; Solnet et al, 2020).

*Designing for Effectiveness*

The "effectiveness" dimension of a content moderator's experience centers on empowering them to accomplish their tasks with precision and fostering a sense of professional competence. From a UX perspective, designing for effectiveness requires a deliberate focus on minimizing cognitive load, ensuring robust usability, and optimizing the critical user journeys inherent in content review. This is important because moderators must make hundreds to thousands of decisions per shift, operating under pressures of stringent accuracy and efficiency metrics that demand both speed and quality (Ahmad & Krzywdzinski, 2022; Barrett, 2020; Steiger et al., 2021).

The challenge is significantly compounded by deficient tooling. Poorly designed systems and tools, plagued by glitches, latency issues, and fragmented ecosystems, directly undermine a moderator's effectiveness. These issues exacerbate cognitive load by forcing frequent context switching, such as needing to open multiple windows to consult policy documents or user history, which fractures concentration and introduces errors. They also promote repetitive manual tasks, like copying and pasting case information, that could easily be automated (Lenaerts & Waeyaert, 2022). In this environment, every unnecessary click and second of load time contributes to mental fatigue and frustration.

Therefore, investing in high-quality tooling is a critical lever for enabling better and more consistent decision-making. This involves designing for effortless interaction by maximizing the reliability and speed of core review tools and creating a unified interface that minimizes the need to switch between different applications. Additionally, effectiveness can be enhanced by leveraging intelligent assistance. Such systems can reduce cognitive load by automating routine data entry, providing relevant contextual information (e.g., a user's recent violation history), or offering timely, guided support by highlighting the most relevant section of a complex policy based on the content being reviewed. Well-designed tools that are reliable, integrated, and intelligent can significantly lower cognitive burden and frustration. This, in turn, protects the wellbeing of the human content moderator and the quality of their work.

### *Designing for Connection*

Connection is one of the most critical protective factors in challenging work environments, acting as a buffer against stress and burnout (Barleycorn, 2022; Newell & MacNeil, 2010; Robertson et al., 2016; Steinlin et al., 2017; Trippany et al., 2004; Vîrgă et al., 2020). This principle is especially true for content moderators, who often work in physical or psychological isolation, making a sense of community and mutual support vital to their wellbeing (Lenaerts & Waeyaert, 2022; Steiger et al., 2021). The connection dimension, therefore, is centered on enabling users to collaborate, communicate, and form meaningful bonds with others (Oeij et al., 2017). User Experience (UX) design can directly support this by embedding features that foster collaboration, facilitate communication, and encourage recognition within the workflow.

At a practical level, this means creating intuitive channels for content moderators to easily seek advice from peers on ambiguous cases, perhaps through dedicated chat functions or a system for flagging a case for a second opinion. It also involves designing clear and seamless pathways to escalate complex issues to subject matter experts or policy specialists, ensuring that no moderator feels they are making a difficult decision entirely alone. Beyond functional collaboration, it is well understood that positive recognition and communication is fundamental to effective teamwork and building trust. Systems that allow content moderators to recognize accomplishments and celebrate each other's successes, such as through peer-to-peer bonuses, team-based awards, or achievement badges, can promote positive relationships and reinforce a culture of mutual respect. Furthermore, features that allow for personalization and self-expression, such as customizable profiles or avatars, can also support interpersonal connections.

Lastly, UX can help connect the moderator's daily tasks to the health of the internet and the impact on end users. This can be achieved by designing features that showcase the real-world value and positive impact of their work. For example, a dashboard could display aggregate data on how many users were protected from harmful content due to the team's efforts, or periodically surface anonymized, positive feedback from the user community. By making the beneficial consequences of their labor visible, such features help support a sense of purpose, countering feelings of futility that can arise from repetitive work (LaMontagne et al., 2014). As a reminder, one of the primary wellbeing goals is to cultivate positive experiences, and celebrating accomplishments and showcasing positive impact are great opportunities to do just that.

### *Designing for Resilience*

Designing for resilience, in the context of content moderation, means holistically addressing emotional, cognitive, and behavioral needs (Caringi et al., 2012; LaMontagne et al., 2014; Miller & Foster, 2010). This involves a two-pronged approach that mitigates the direct harm of content exposure and proactively promotes recovery.

**Mitigating Direct Content Exposure.** When moderators must engage with harmful material, "content soothing" features can serve as a line of defense. These tools, which include options like blurring, applying a grayscale filter, or muting audio by default, are designed to reduce the immediate sensory intensity of potentially egregious content (Ahmad & Krzywdzinski, 2022; Barrett, 2020; Dang et al., 2018; Karunakaran & Ramakrishan, 2019; Steiger et al., 2021). By diminishing the realism of the media, these features help create emotional distance between the moderator and their work, lessening the potential for psychological harm. However, the implementation of these tools is critical. They must be optional and well-built; if content soothing features are clumsy, slow, or negatively impact a moderator's ability to perform their tasks quickly and accurately, they are unlikely to be used. Worse, they can become a source of frustration, which defeats their intended wellbeing benefits.

**Promoting Recovery Through Thoughtful Interruption.** Sustaining accuracy and speed in a cognitively demanding role like content moderation relies on robust yet limited cognitive resources that deplete without adequate recovery. It can also be difficult for an individual to self-monitor their own emotional state while simultaneously engaging in intense cognitive tasks. Therefore, UX design should aim to thoughtfully interrupt excessive "flow," especially when a moderator is continuously exposed to sensitive content. This can be achieved with just-in-time interventions, such as automated reminders to take breaks that are triggered by exposure duration or content severity (Steiger et al., 2021).

An effective take-a-break feature can help initiate the recovery process without placing additional demands on the self-monitoring necessary for purely self-initiated breaks. The benefits of breaks are well-documented across multiple disciplines, grounded in theoretical foundations like the Conservation of Resources (COR) theory, the Effort-Recovery Model (ERM), Attention Restoration Theory (ART), and Stress-Recovery Theory (SRT; Albulescu et al., 2022; Hobfoll, 1989; Kaplan, 1995; Meijman & Mulder, 2013). This body of research highlights that the ideal break structure is highly dependent on the nature of the work. Complex or emotionally taxing tasks may require longer recovery periods of 20 minutes or more to be effective. In contrast, highly repetitive or visually demanding work often benefits from frequent, short breaks away from screens, which can help prevent eye strain and mental fatigue (Albulescu et al., 2022; Lee et al., 2015).

To be successful, these break systems must promote autonomy. Moderators should be able to customize their settings, including the frequency, duration, and type of activity (Manganelli et al., 2018; Slemp et al., 2015). A well-designed system might offer a menu of evidence-based suggestions, such as short physical activities to engage the body, mindfulness exercises to calm the nervous system, and visuospatial games to minimize the formation of intrusive memories following exposure to egregious content (Henning et al., 1997; Holmes et al., 2010; Kessler et al., 2020; Lacaze et al., 2010; Zacher et al., 2014).

**Beyond Features: Integrating Wellbeing into UX Research, Writing, and Design**

To effectively design for wellbeing, it must be treated as a core principle woven throughout the entire design cycle, not as an add-on feature. This approach begins with the recognition that every design choice, even those not explicitly intended for wellbeing, can impact a user's state of being. Consequently, designing with the whole person and their complete experience in mind becomes essential. A key step in this process is mapping the user journey, which allows researchers and designers to proactively identify specific "wellbeing touchpoints." These are critical moments where targeted interventions can prevent harm, mitigate negative experiences, or foster positive ones.

*UX Research*

During the research phase, a process which aims to understand user behaviors, needs, and motivations through observation and feedback, we encourage three important considerations to effectively design for wellbeing.

First, it is crucial to identify the person's broader needs and pain points. Frameworks like the Critical User Journey (CUJ) are invaluable here. A CUJ maps out the entire sequence of steps a user takes to accomplish a specific, high-value goal, allowing teams to analyze the entire experience from start to finish. By mapping these journeys (e.g., "reviewing a piece of sensitive content"), researchers can pinpoint specific "wellbeing touchpoints" where negative experiences can be mitigated or positive ones can be introduced. While other frameworks exist, such as Jobs to be Done (JTBD), which focuses on the underlying "job" a user is trying to accomplish, the CUJ is particularly effective at visualizing the emotional and cognitive highs and lows of a process, making it well suited for identifying where wellbeing levers are most needed.

Second, because wellbeing is a multi-dimensional concept with no single metric capturing its complexity, it is essential to collect both subjective and objective data (Cooke et al., 2016; Miller & Foster, 2010; Wassell & Dodge, 2015). Subjective data, or what people say, can be gathered through surveys, interviews, and self-reported mood ratings to understand their perceptions and feelings. Objective data, or what they do, includes behavioral metrics like task completion times, error rates, or tool usage patterns. Relying on only one type of signal can be misleading; for example, a user might report feeling fine (subjective) while their performance metrics show a clear decline (objective). A robust measurement program must integrate multiple signals, including standard UX metrics, emotional signals (e.g., a periodic mood pulse survey), cognitive signals (e.g., attentional capacity or beliefs), and behavioral signals (e.g., productivity and errors).

Third, researchers must be particularly mindful of ethical considerations. Wellbeing data is inherently sensitive, as it touches upon an individual's mental, emotional, and sometimes physical state. There are often specific legal and ethical protections around collecting and storing anything that could be construed as healthcare data. If you are collecting any healthcare data, ensure that your process is compliant with data privacy regulations like HIPAA (Health Insurance Portability and Accountability Act) in the US or analogous regulations in the region where research is conducted. Additionally, if you are collecting data from commercial content moderators, follow the employment or co-employment policies within your organization to ensure compliance with government labor laws and regulations. Lastly, always obtain clear, informed consent and build trust with participants by being transparent about how their information will be used to improve their experience, not to evaluate their performance.

While the direct collection of biometric or neurophysiological data by a platform is typically impractical and poses significant ethical challenges, there is potential in empowering users to leverage their own data. For instance, encouraging moderators to pay attention to data from their personal devices (e.g., smartwatches that track heart rate or stress levels) can be a helpful tool for self-monitoring. This approach reframes data collection not as a corporate mandate, but as a personal resource, helping individuals recognize their own emotional state and intervene accordingly for their own benefit.

*__UX Writing__*

Effective communication is critical when designing for wellbeing, and this is the domain of UX Writing and Content Strategy. UX Writing is the practice of crafting the copy that appears directly within a user interface: menus, error messages, and instructional prompts that guide users and help them accomplish their goals. Content Strategy is the higher-level discipline that governs all of the product's content, ensuring it is useful, usable, and well structured. This includes creating the information architecture (IA), which is the blueprint for how information is organized and labeled. In contexts that require users to consume large amounts of complex information, such as a policy center, a well-defined content strategy and clear UX writing are critical for preventing cognitive overload and ensuring comprehension.

When applied to the topic of wellbeing, these practices require special consideration, as wellbeing can be a "sticky subject" with ambiguous and highly personal terminology (Cooke et al., 2016; Miller & Foster, 2010; Needs, 2010; Wood, 2020). The tone and voice should be personable yet professional, consistently inclusive, and free from judgment. Designing features to promote wellbeing requires careful attention to terminology, given the topic's sensitive and nuanced nature. This includes avoiding "fuzzy" or unclear wellness terms that are common in some wellness or influencer industries, but lack evidence-based support.

Furthermore, UX writing must account for the heterogeneous experiences of content moderators, avoiding language that presumes all individuals are affected in the same way (Ahmad & Krzywdzinski, 2022; Bryant, 2001; Richardson & Rothstein, 2008). The copy should never place blame but instead empower moderators to feel that they can take action. For example, instead of a label that says, "Feeling overwhelmed?" which presumes a personal struggle, a more empowering and proactive option might be, "Manage your workload" or "Adjust content settings."

It is also important to be precise and responsible with clinical language. Terms for diagnosable mental health conditions, such as post-traumatic stress disorder (PTSD), must never be used in a non-clinical sense or interchangeably with general terms like stress (Barleycorn, 2022; Newell & MacNeil, 2010; Trippany et al., 2004). A feature designed to offer immediate help should not be labeled "Crisis Support," which can be alarming and stigmatizing. A more carefully considered UX writing choice would be "Get Immediate Support" or "Talk to Someone Now," which is actionable, less clinical, and focuses on the solution.

Finally, a global content strategy must recognize the significant stigma associated with mental health in many cultures, which can influence how individuals perceive and pursue wellbeing (LaMontagne et al., 2014; Steinlin et al., 2017; Tay & Diener, 2011). It is essential to rely on culturally sensitive research to adapt language accordingly, while always acknowledging the inherent heterogeneity of experiences within any cultural background.

***UX Design***

Designing for the wellbeing of content moderators is a thoughtful application of core UX principles, requiring a deliberate effort to minimize cognitive load, ensure usability, and optimize user journeys, particularly within the often "scrappy" context of internal tool development. As previously described, this effort is guided by key principles: designing for effectiveness, connection, and resilience by prioritizing effortless and intelligent design; actively creating emotional distance from sensitive content; supporting recovery through structured breaks; fostering community and collaboration; and ensuring all interventions are unified by a commitment to user autonomy (Bharucha et al., 2022; Slemp et al., 2015; Steiger et al., 2021; Wassell & Dodge, 2015). To operationalize these principles, design and ideation processes need to incorporate specific, proactive methodologies that embed wellbeing as a foundational requirement.

**Adversarial Design Workshops.** A primary methodology is the adoption of adversarial design workshops, a practice adapted from security "red teaming" and critical design theory (DiSalvo, 2012). In these sessions, a design team's objective is to intentionally conceptualize the worst possible experience for moderator wellbeing. By attempting to maximize stress, cognitive load, or frustration, they can uncover latent risks and anti-patterns that would not be apparent during standard, "positive path" design exercises. This approach is a form of Failure Mode and Effects Analysis (FMEA), systematically identifying how a system might fail its user not just functionally, but psychologically (Stamatis, 2003). The output is a clearer understanding of potential harms that must be designed against, ensuring a more resilient and protective final product.

**Participatory Design.** Second, teams must move beyond simple user feedback to participatory design, or co-design, methodologies. This approach positions end-users, in this case, content moderators, as active expert collaborators in the design process itself (Sanders & Stappers, 2008; Schuler & Namioka, 1993). Rather than merely validating concepts, moderators participate in ideation, sketching, and workflow mapping, contributing their deep, lived expertise. For high-risk, specialized roles, this method is helpful for ensuring that wellbeing features are not only well-intentioned but are also practical and genuinely empowering within the established workflow (Robertson & Simonsen, 2012). Co-design directly supports the principle of user autonomy, as the tools are shaped by the people who will ultimately use and control them.

**Wellbeing Checklists.** Finally, wellbeing principles must be formalized within established development frameworks by creating wellbeing-centric design criteria. This involves integrating a "Wellbeing Checklist" or similar heuristic evaluation tool into a team's "Definition of Done" for any new feature. Drawing from disciplines like Value Sensitive Design, which advocates for proactively incorporating human values into technology, this checklist provides a concrete mechanism for accountability (Friedman & Hendry, 2019). It prompts designers and developers to explicitly consider and document factors such as cognitive load, potential for emotional harm, and support for user autonomy before a feature can be approved. This formalization shifts wellbeing from an abstract ideal to a tangible product requirement.

**A Holistic Approach**

Content moderation is a demanding role, characterized by repetition, high-stakes decisions, and ongoing exposure to harmful material. Therefore, protecting content moderators requires a deliberate, evidence-based approach where wellbeing is a core principle woven into the product design lifecycle for the tools they use. User experience (UX) and tooling have an important role to play in supporting the wellbeing of content moderators, moving beyond reactive measures to a holistic strategy encompassing prevention, timely intervention, and the cultivation of positive experiences.

A holistic approach is important because it accounts for unintended consequences, which is especially relevant for content moderation where well intentioned features may backfire by having a negative impact on one of the interdependent core dimensions of wellbeing. When designing for effectiveness, connection, and resilience, it is helpful to take a step back and consider how each feature as well as the entire design may impact content moderators' emotion, cognition, and behavior across all three dimensions. Without this careful consideration, even seemingly beneficial designs can inadvertently create new challenges for moderators.

Wellbeing features should be grounded in evidence-based strategies, drawing valuable insights from adjacent literatures. For example, clinical and health psychology offer a wealth of research on stress reduction and behavioral change. Strategies from these fields can inform wellbeing features, which should then be tested to ensure their efficacy and generalizability to the context of content moderation user experience. In addition to being evidence based, these features must be built with high quality; if they are clunky, buggy, or suffer from poor usability, they will introduce friction. All these design elements must be implemented in a way that empowers content moderator autonomy, giving them control over tools and fostering a sense of competence and confidence.

Bringing these technological and feature-level solutions to life requires a holistic perspective, viewing tooling as inseparable from the human and social elements of the work. This aligns with established concepts like Sociotechnical Systems Design (STS-D), which emphasizes the joint optimization of technical and social conditions. By embedding wellbeing focused research, writing, and design practices into the development cycle, organizations can move beyond a reactive or "scrappy" approach to internal tooling. Investing in this comprehensive, user-centered process is not merely an ethical obligation; it is fundamental to the long-term viability and success of building healthier, more sustainable, and ultimately more effective trust and safety operations for the benefit of all users.

**Conclusion**

Content moderation, while vital for safe online environments, places immense and often hidden demands on human moderators. This chapter has advocated for a wellbeing-centered approach as an essential strategy. We've shown how UX can move beyond reactive measures to a holistic approach of prevention, timely intervention, and the cultivation of positive experiences. This means designing for effectiveness with intuitive tools that reduce cognitive load, fostering connection through collaboration and visible impact, and building resilience via content soothing features and restorative breaks.

Integrating wellbeing into UX fundamentally requires it to be a core principle across research, writing, and design. This involves empathetic research grounded in evidence-based literature, clear and sensitive UX writing, and proactive design methodologies like adversarial workshops and participatory design. The key argument presented is that prioritizing moderator wellbeing through thoughtful UX is a strategic investment that leads to more sustainable and effective trust and safety operations, benefiting all online users and businesses.

**Limitations and Future Directions**

While this chapter offers a comprehensive framework, its limitations include the nascent state of direct research on moderator wellbeing and the challenges of implementing these strategies in resource-constrained settings. Future efforts must focus on longitudinal studies and scalable integration solutions.

Looking ahead, generative AI is set to transform content moderation by shifting the human role toward handling the most ambiguous, ethically complex, and high-risk cases. This transformation also introduces additional human roles working on the design, deployment, and oversight of automated systems. These individuals may be exposed to sensitive or egregious content during model development or monitoring, posing challenges to wellbeing. UX design must evolve to support effective human-AI collaboration, including effectiveness through calibrated trust and resilience through evidence-based support. A proactive UX approach will ensure human wellbeing remains central in an increasingly AI-driven moderation landscape.